\documentclass[prd,twocolumn,showpacs,preprintnumbers,amsmath,amssymb]{revtex4}

\usepackage{graphics,color}
\usepackage{epsfig}
\usepackage{dcolumn}
\usepackage{bm}

\begin{document}
\title{The signal of $Z^\pm(4430)$ in nucleon-antinucleon scattering}

\author{Hong-Wei Ke$^{1}$}
\author{Xiang Liu$^{2}$\footnote{Corresponding author}}\email{liuxiang@teor.fis.uc.pt}

\affiliation{$^1$Department of Physics, Nankai University, Tianjin
300071, China\\
$^2$Centro de F\'{i}sica Computacional, Departamento de
F\'{i}sica, Universidade de Coimbra, P-3004-516 Coimbra, Portugal}

\date{\today}

\begin{abstract}

We study the production of $Z^\pm(4430)$ at a nucleon-antinucleon
scattering experiment. Considering the PANDA experiment to be an
ideal platform to explore the production of the charmonium and
charmonim-like states, we suggest the forthcoming PANDA experiment
to pay attention to the production of $Z^\pm(4430)$.
\end{abstract}


\maketitle

\section{introduction}\label{sec1}

Recently the Belle Collaboration announced the observation of
$Z^\pm(4430)$ in the invariant mass spectrum of $\psi'\pi^\pm$ of
$B\to K\psi'\pi^\pm$. As an excellent candidate of the exotic
state, $Z^\pm(4430)$ possesses
$m=4433\pm4(\mathrm{stat})\pm1(\mathrm{syst})$ MeV and
$\Gamma=44^{+17}_{-13}(\mathrm{stat})^{+30}_{-11}(\mathrm{syst})$
MeV \cite{Belle-4430}. Up to now, theorists have paid extensive
attention to $Z^{\pm}(4430)$
\cite{rosner,Meng,xiangliu,xiangliu-2,ding,qsr,maiani,cky,Gershtein,Bugg2,Qiao,braaten,lu,close,open,Matsuki}.
$Z^\pm(4430)$ was explained as an S-wave threshold effect
\cite{rosner}, a $D_1\bar D^*/D_1'\bar D^*$ molecular state
\cite{Meng,xiangliu,xiangliu-2,ding,qsr,lu}, a tetraquark state
\cite{maiani,cky,Gershtein,close}, a cusp effect \cite{Bugg2}, a
$\Lambda_c-\bar\Sigma_c^0$ bound state \cite{Qiao} and the
radially excited state of $D_s$ \cite{Matsuki}. Besides exploring
the structure of $Z^\pm(4430)$, the hidden and open charm decays
were studied in Ref. \cite{Meng,open}.

Up to now, $Z^\pm(4430)$ was only observed by the Belle
experiment. Since searching $Z^\pm(4430)$ in other experiments
will be helpful to establish $Z^\pm(4430)$ finally, the
theoretical study of the production of $Z^\pm(4430)$ is an
interesting topic. For example, recently the author of Ref.
\cite{close} studied the production of $Z^\pm(4430)$ by the
photoproduction experiment.

Exploring the charmonium and charmonium-like states above the open
charm threshold is one of the most important physics aims of the
forthcoming PANDA (AntiProton Annihilations at Darmstadt)
experiment at FAIR \cite{panda}. Recently some groups performed
the phenomenological studies of the productions of the charmonium
and charmonium-like states around the PANDA experiment. Barnes and
Li calculated the differential and total cross sections of the
near-threshold associated charmonium production processes
$p\bar{p}\to \pi\Psi$ with
$\Psi=\eta_c,J/\psi,\psi',\chi_{c0},\chi_{c1}$ using a simple
initial-state light meson emission model \cite{barnes}. In Ref.
\cite{ma}, the authors studied the production of $X(3872)$ at the
PANDA experiment assuming $X(3872)$ to be the molecular state
$D\bar D^*$ or the charmonium state $\chi_{c1}(2P)$. These
instructive explorations not only further guide the PANDA
experiment to find charmonium or charmonium-like state but also
stimulate more theorists to focus on this field.

The PANDA experiment could be an ideal platform to explore the
production of $Z^\pm(4430)$. Thus, in this work, we first carry
out a phenomenological study of $Z^+(4430)$ production at
antiproton-proton scattering experiment. Meanwhile we also
tentatively investigate $Z^-(4430)$ production by the
antiproton-neutron annihilation process.

This work is organized as follows. After the introduction, we
present the formula of the production of $Z^+(4430)$ by the
$p\bar{p}$ interaction. Next, we formulates the expressions of
$Z^-(4430)$ production by the $\bar p n$ annihilation in Sec.
\ref{sec2}. Section \ref{sec4} is for numerical results and our
discussion.

\section{The Production of $Z^+(4430)$ by the scattering of proton and antiproton}\label{sec1}

In 1987, Gaillard, Maiani and Petronzio for the first time applied
the hadron-level pole diagram to calculate the production of
charmonium from the $p\bar{p}$ states with fixed quantum number
\cite{GMP}. Later Lundborg, Barnes and Wiedner developed this
method to do in-depth study of $p\bar{p}\to \pi^0\Psi\;
(\Psi=\eta_c,J/\psi,\psi',\chi_{c0},\chi_{c1})$ \cite{barnes,LBW}.
The hadron-level pole diagram is a suitable approach to study the
charmonium or charmonium-like state in the low-energy $p\bar p$
scattering process.

\begin{figure}[htb]
\begin{center}
\begin{tabular}{ccccccc}
\scalebox{0.6}{\includegraphics{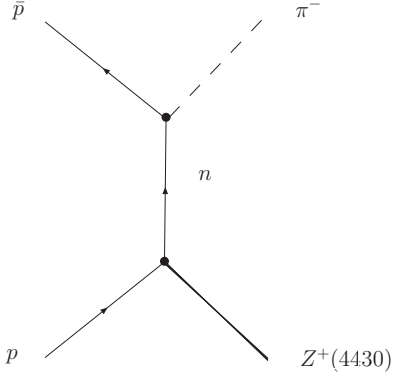}}
\end{tabular}
\end{center}
\caption{The diagram depicting the production of $Z^+(4430)$ in
the proton-antiproton scattering process. \label{production}}
\end{figure}

As an excellent candidate for the exotic state, the component of
$Z^{+}(4430)$ should be $c\bar{c}q_1\bar{q}_2$ for explanations in
terms of the $D_{1}\bar D^*$ molecular state
\cite{xiangliu,xiangliu-2} or the tetraquark state \cite{maiani}.
Thus the $Z^+(4430)$ production mechanism for the $p\bar p$
scattering process is similar to $p\bar{p}\to \pi^0\Psi$. In this
work, we use the so called hadron-level pole diagram to depict the
production of $Z^+(4430)$ at the proton and antiproton scattering
process, which is depicted in Fig. \ref{production}. By exchanging
neutron between proton and antiproton, the process of $p\bar{p}\to
Z^{+}(4430)\pi^-$ occurs.

In this work, we consider three possible quantum numbers for
$Z^+(4430)$, i.e. $J^{P}=0^-,1^-,2^-$ \cite{xiangliu}. The
effective Lagrangians describing the interaction of $Z^+(4430)$
with nucleons are written as
\begin{eqnarray}
\mathcal{L}_{{Z^+p\bar n}} &=& \left\{\begin{array}{lcc}
g_{_{Zpn}}^{(0)}\bar{n}\gamma_5pZ & \mbox{for} & J^{P}=0^-\\
ig_{_{Zpn}}^{(1)}\bar{n}\gamma_{\mu}pZ^{\mu} & \mbox{for} & J^{P}=1^-\\
g_{_{Zpn}}^{(2)}\gamma_5\partial_{\mu}\bar{n}\partial_{\nu}pZ^{\mu\nu} & \mbox{for} &J^{P}= 2^-\\
\end{array}\right .\label{2}
\end{eqnarray}
The value of the coupling constants $g_{_{Zpn}}^{(0),(1),(2)}$
will be given in the section on the numerical results. The
Lagrangian relevant to the coupling of $\pi$ and the nucleons is
well established \cite{coupling-nnpi}:
\begin{eqnarray}
&&\mathcal{L}_{\bar NN\pi} =-i\mathcal{G}\bar{N}\gamma_5{\mbox{\boldmath $\tau$}}\cdot
 {\mbox{\boldmath $\pi$}}N \nonumber\\
&&\quad=-i \sqrt 2\mathcal{G}\bigg[\bar p\gamma_5n\pi^++\bar
n\gamma_5p\pi^-+\frac{(\bar p\gamma_5p-\bar n\gamma_5
n)\pi^0}{\sqrt{2}}\bigg],\nonumber\\
\end{eqnarray}
where $\bar N=(\bar p,\bar n)$. The coupling constant
$\mathcal{G}$ is taken to be $13.5$.

Using the above Lagrangian, we write out the amplitudes of the
$p\bar{p}\to Z^+(4430)\pi^-$ process depicted in Fig.
\ref{production}
\begin{eqnarray}
\mathcal{M}^{J=0}&=&\sqrt{2}
\mathcal{G}g^{(0)}_{_{pnZ}}\bar{\nu}_{_{\bar{p}}}\gamma_5\frac{1}
{p\!\!\!\slash_2-p\!\!\!\slash_b-m_n }\gamma_5
u_p,\\
\mathcal{M}^{J=1}&=&\sqrt{2}i
\mathcal{G}g^{(1)}_{_{pnZ}}\bar{\nu}_{_{\bar{p}}}\gamma_5
\frac{1}{p\!\!\!\slash_2-p\!\!\!\slash_b-m_n }\gamma_\mu
u_p\varepsilon^\mu,\\
\mathcal{M}^{J=2}&=&-\sqrt{2}
\mathcal{G}g^{(2)}_{_{pnZ}}\bar{\nu}_{_{\bar{p}}}\gamma_5 \frac{1}
{p\!\!\!\slash_2-p\!\!\!\slash_b-m_n }\gamma_5 u_p
{p_a}^\mu{p_b}^\nu \varepsilon_{\mu\nu},\nonumber\\
\end{eqnarray}
respectively, corresponding to the production of $Z^+(4430)$ with
$J^{P}=0^-,1^-,2^-$. Here $p_a$, $p_b$, $p_1$ and $p_2$ denote the
four-momentum of $p$, $\bar{p}$, $Z^+(4430)$ and $\pi^-$
respectively.

\begin{figure}[htb]
\begin{center}
\begin{tabular}{ccccccc}
\scalebox{0.7}{\includegraphics{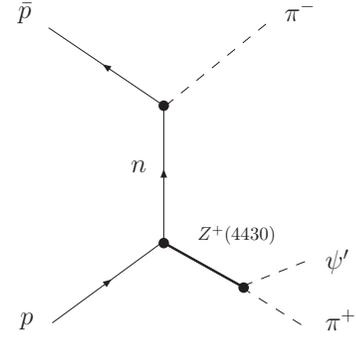}}
\end{tabular}
\end{center}
\caption{The diagram depicts $p\bar p\to \pi^-Z^+(4430)\to
\pi^-\psi'\pi^+$. \label{production-1}}
\end{figure}

Since $Z^{+}(4430)$ is observed in the invariant mass spectrum of
$\psi'\pi^+$; thus, we further explore $p\bar{p}\to
Z^{+}(4430)\pi^-\to \psi'\pi^+\pi^-$ process, depicted in Fig.
\ref{production-1}.

The interactions of $Z^+(4430)$ with $\pi^-$ and $\psi'$ are
\begin{eqnarray}
&&\mathcal{L}_{{Z^+\pi^-\psi'}} \nonumber\\&&=
\left\{\begin{array}{lcc}
ig_{_{Z\pi\psi'}}^{(0)}(\pi^-\partial_{\mu}Z-\partial_{\mu}\pi^-Z)\psi^{\prime\mu}&\mbox{for}& J^{P}=0^-\\
ig_{_{Z\pi\psi'}}^{(1)}\varepsilon_{\mu\nu\alpha\beta}\partial^{\mu}\psi^{\prime\nu}\partial^{\alpha}
Z^{\beta}\pi^-+h.c.&\mbox{for}& J^{P}=1^-\\
ig_{_{Z\pi\psi'}}^{(2)}\psi^{\prime}_{\mu}Z^{\mu\nu}\partial_{\nu}\pi^-&\mbox{for}&J^{P}= 2^-\\
\end{array}\right.\nonumber\\\label{3}
\end{eqnarray}
The coupling constants $g_{_{Z\pi\psi'}}^{(0),(1),(2)}$ will be
discussed in Sec. \ref{sec4}. It is straightforward to obtain the
amplitudes:
\begin{eqnarray}
\mathcal{M}^{J=0}&=&\sqrt{2}
\mathcal{G}g^{(0)}_{_{pnZ}}g^{(0)}_{_{Z\pi^+\psi'}}\bar{\nu}_{_{\bar{p}}}\gamma_5\frac{1}
{p\!\!\!\slash_3-p\!\!\!\slash_b-m_n }\gamma_5
u_p\nonumber\\&&\times\frac{1}{(q^2-{m_Z}^2+im_Z\Gamma_Z)}(p_1+2p_2)\cdot
\varepsilon ,\nonumber\\\label{2-3-1}
\end{eqnarray}

\begin{eqnarray}
\mathcal{M}^{J=1}&=&\sqrt{2}i
\mathcal{G}g^{(1)}_{_{pnZ}}g^{(1)}_{_{Z\pi^+\psi'}}\bar{\nu}_{_{\bar{p}}}\gamma_5
\frac{1}{p\!\!\!\slash_3-p\!\!\!\slash_b-m_n }\gamma_\nu
u_p\nonumber\\&&\times\frac{g_{\nu\beta}-q_\nu
q_\beta/q^2}{(q^2-{m_Z}^2+im_Z\Gamma_Z)}\epsilon_{\rho\mu\alpha\beta}p_1^\rho\varepsilon^\mu(p_1+p_2)^\alpha,\nonumber\\
\label{2-3-2}
\end{eqnarray}

\begin{eqnarray}
\mathcal{M}^{J=2}&=&-\sqrt{2}
\mathcal{G}g^{(2)}_{_{pnZ}}g^{(2)}_{_{Z\pi^+\psi'}}\bar{\nu}_{_{\bar{p}}}\gamma_5
\frac{1} {p\!\!\!\slash_3-p\!\!\!\slash_b-m_n }\gamma_5
u_p\nonumber\\&&\times \frac{1}
{(q^2-{m_Z}^2+im_Z\Gamma_Z)}{p_a}^\alpha{(p_3-p_b)}^\beta
\varepsilon^\mu {p_2}^\nu\nonumber\\
&&\times\left(\frac{q_{\mu\alpha}q_{\nu\beta}+q_{\mu\beta}q_{\nu\alpha}}{2}-\frac{q_{\mu\nu}q_{\alpha\beta}}{3}
\right) \label{2-3-3}
\end{eqnarray}
respectively corresponding to the production of $Z^+(4430)$ with
$J^{P}=0^-,1^-,2^-$. Here $q=q_1+q_2$ and $q_{ab}=-g_{ab}+q_a
q_b/q^2$.

The cross section of $p\bar{p}\to \psi'\pi^+\pi^-$ with
intermediate state $Z^+(4430)$, which involves three final states
$\psi'$, $\pi^+$ and $\pi^-$, can be obtained by integrating over
the phase space \cite{3-body}
\begin{eqnarray}
&&\sigma(p\bar{p}\rightarrow
\psi'\,\,\pi^+\,\,\pi^-)\nonumber\\&&=\int^{a_{2}}_{a_{1}}
\mathrm{d}p_2^0\int^{b_{2}}_{b_{1}}
\mathrm{d}p_{3}^0\int^{2\pi}_{0} \mathrm{d}\eta
\int^{1}_{-1}\mathrm{d}(\cos\theta) 
d\sigma
,\nonumber\\\label{NI}
\end{eqnarray}
where $a_1$, $a_2$, $b_{1}$ and $b_2$ are defined as respectively
\begin{eqnarray*}
a_{1}&=&m_2,\;\;\; a_{2}=\frac{{\sqrt{s}}}{2}- \frac{ {( {m_{1}} +
{m_{3}} ) }^2-m_2^2}
   {2\,{\sqrt{s}}},\\
b_{1}&=&\frac{1}{2\,\tau}\left[\rho(\tau+m_{+}m_{-})-|\mathbf{p}_2|\sqrt{(\tau-m_{+}^2)(\tau-m_{-}^2)}\right],\\
b_{2}&=&\frac{1}{2\,\tau}\left[\rho(\tau+m_{+}m_{-})+|\mathbf{p}_2|\sqrt{(\tau-m_{+}^2)(\tau-m_{-}^2)}\right],\\
\rho&=&\sqrt{s}-p_2^0,\;\;\; \tau=\rho^2-|\mathbf{p}_2|^2,\;\;\;
m_{\pm}=m_{3}\pm m_{1}.
\end{eqnarray*}
The definition of $d\sigma$ reads as
\begin{eqnarray*}
\\d\sigma&=&\frac{1}{32(2\pi)^4|\mathbf{p}_a|\sqrt{s}}|\mathcal{M}|^2
.
\end{eqnarray*}
Here $m_{1,2,3}(p_{1,2,3})$ denote the masses (four-momentum) of
$\psi'$, $\pi^+$ and $\pi^-$, respectively. $\mathcal{M}$
represents the amplitude shown in (\ref{2-3-1}-\ref{2-3-3}).

\section{The Production of $Z^-(4430)$ in the antiproton-neutron annihilation process}\label{sec2}

\begin{figure}[htb]
\begin{center}
\begin{tabular}{ccccccc}
\scalebox{0.6}{\includegraphics{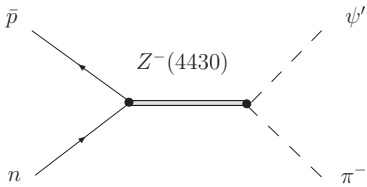}}
\end{tabular}
\end{center}
\caption{The production of $Z^+(4430)$ by antiproton-neutron
annihilation. \label{PNZ}}
\end{figure}

\begin{figure}[htb]
\begin{center}
\begin{tabular}{ccccccc}
\scalebox{0.5}{\includegraphics{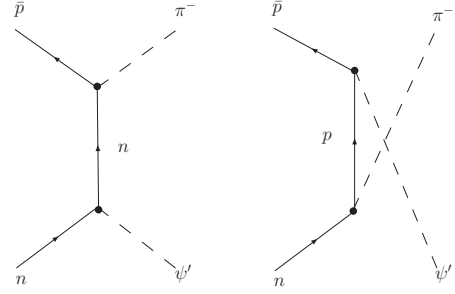}}
\end{tabular}
\end{center}
\caption{The background of the production of $Z^+(4430)$ in the
antiproton-neutron interaction process. \label{APN}}
\end{figure}

In this section, we study the process that $\bar{p}n\to
Z^-(4430)\to\psi'+\pi^-$, which is depicted in Fig. \ref{PNZ}.
Figure \ref{APN} describes the background $\bar p n\to \psi'\pi^-$
relevant to the production of $Z^+(4430)$.

Thus, we can directly obtain the amplitudes for the production of
$Z^+(4430)$ at antiproton-neutron annihilation, which are
represented as
\begin{eqnarray}
\mathcal{M}^{J=0}&=&
g^{(0)}_{_{Zpn}}g^{(0)}_{_{Z\pi\psi'}}\bar{\nu}_{_{\bar{p}}}\gamma_5
u_n\nonumber\\&&\times\frac{1}{(q^2-{m_Z}^2+im_Z\Gamma_Z)}(p_1+2p_2)\cdot
\varepsilon ,\nonumber\\
\end{eqnarray}

\begin{eqnarray}
\mathcal{M}^{J=1}&=&i
g^{(1)}_{_{Zpn}}g^{(1)}_{_{Z\pi^+\psi'}}\bar{\nu}_{_{\bar{p}}}\gamma_\nu
u_n\nonumber\\&&\times\frac{g_{\nu\beta}-q_\nu
q_\beta/q^2}{(q^2-{m_Z}^2+im_Z\Gamma_Z)}\epsilon_{\rho\mu\alpha\beta}p_1^\rho\varepsilon^\mu(p_1+p_2)^\alpha,\nonumber\\
\end{eqnarray}

\begin{eqnarray}
\mathcal{M}^{J=2}&=&-g^{(2)}_{_{Zpn}}g^{(2)}_{_{Z\pi^+\psi'}}\bar{\nu}_{_{\bar{p}}}\gamma_5
u_n\nonumber\\&&\times \frac{1}
{(q^2-{m_Z}^2+im_Z\Gamma_Z)}{p_a}^\alpha{p_b}^\beta
\varepsilon^\mu {p_2}^\nu\nonumber\\
&&\times\left(\frac{q_{\mu\alpha}q_{\nu\beta}+q_{\mu\beta}q_{\nu\alpha}}{2}-\frac{q_{\mu\nu}q_{\alpha\beta}}{3}
\right)
\end{eqnarray}
respectively corresponding to the production of $Z^+(4430)$ with
$J^{P}=0^-,1^-,2^-$. Here $q=q_1+q_2$ and
$q_{\mu\nu}=-g_{\mu\nu}+q_\mu q_\nu/q^2$.

For obtaining the amplitude of the background $\bar p n\to
\psi'\pi^-$ shown in Fig. \ref{APN}, we need the effective
Lagrangian of the strong interaction between $\psi'$ and the
nucleons
\begin{eqnarray}
\mathcal{L}_{\bar{N}N\psi'} &=&ig_{p\bar
p\psi'}p\gamma^{\mu}\bar{p}\psi'_{\mu}+ig_{n\bar
n\psi'}n\gamma^{\mu}\bar{n}\psi'_{\mu}.\label{1}
\end{eqnarray}
Since the partial decay width of $\psi'\to p\bar{p}$ is given
experimentally, the coupling constant $g_{p\bar p\psi'}$ can be
obtained. In Ref. \cite{barnes}, Barnes and Li obtained $g_{p\bar
p\psi'}=(0.97\pm 0.04)\times 10^{-3}$ by the experimental
branching ratio $B(\psi'\to p\bar{p})=(2.65\pm 0.22)\times
10^{-4}$. Due to SU(2) symmetry, one can estimate the relation of
$g_{n\bar n\psi'}$ and $g_{p\bar p\psi'}$ by $g_{n\bar
n\psi'}\approx g_{p\bar p\psi'}$.

The amplitudes of the process $\bar p n\to \psi'\pi^-$ read as
\begin{eqnarray}
\mathcal{M}_a&=&\sqrt{2}i
\mathcal{G}g_{_{nn\psi'}}\bar{\nu}_{_{\bar{p}}}\gamma_5\frac{1}{p\!\!\!\slash_a-p\!\!\!\slash_2-m_n
} \varepsilon\!\!\!\slash u_n ,
\end{eqnarray}
\begin{eqnarray}
\mathcal{M}_b&=&\sqrt{2}i
\mathcal{G}g_{_{pp\psi'}}\bar{\nu}_{_{\bar{p}}}\varepsilon\!\!\!\slash\frac{1}{p\!\!\!\slash_a-p\!\!\!\slash_1-m_p
} \gamma_5 u_n .
\end{eqnarray}

\section{Numerical result and discussion}\label{sec4}

The input parameters include $m_{\psi'}=3686.1$ MeV,
$m_{\pi^{\pm}}=139.6$ MeV, $m_{p}=938.3$ MeV, $m_{n}=939.6$ MeV
\cite{PDG}; $m_{Z}=4433$ MeV, $\Gamma_{Z}=44$ MeV
\cite{Belle-4430}.

In \cite{Meng}, Meng and Chao calculated the decay width of
$Z^{+}(4430)\to \psi'\pi^+$ by rescattering mechanism assuming
$Z^{+}(4430)$ to be an S-wave $D_{1}\bar D^*$ molecular state, and
their result indicates that the order of magnitude of
$\Gamma_{Z\psi'\pi^+}$ is several MeV, which can be applied to
estimate the coupling constants $g_{_{Z\pi\psi'}}^{(0,1,2)}$.
Thus, in this work we adopt the below coupling constants
$$g_{_{Z\pi\psi'}}^{(0)}=0.08,\quad g_{_{Z\pi\psi'}}^{(1)}=
0.05\;\;\mathrm{GeV}^{-1},\quad g_{_{Z\pi\psi'}}^{(2)}=0.33,$$
which are determined by taking $\Gamma_{Z\psi'\pi^+}=1$ MeV.

\begin{widetext}
\begin{center}
\begin{figure}[htb]
\begin{tabular}{ccccccc}
\scalebox{0.55}{\includegraphics{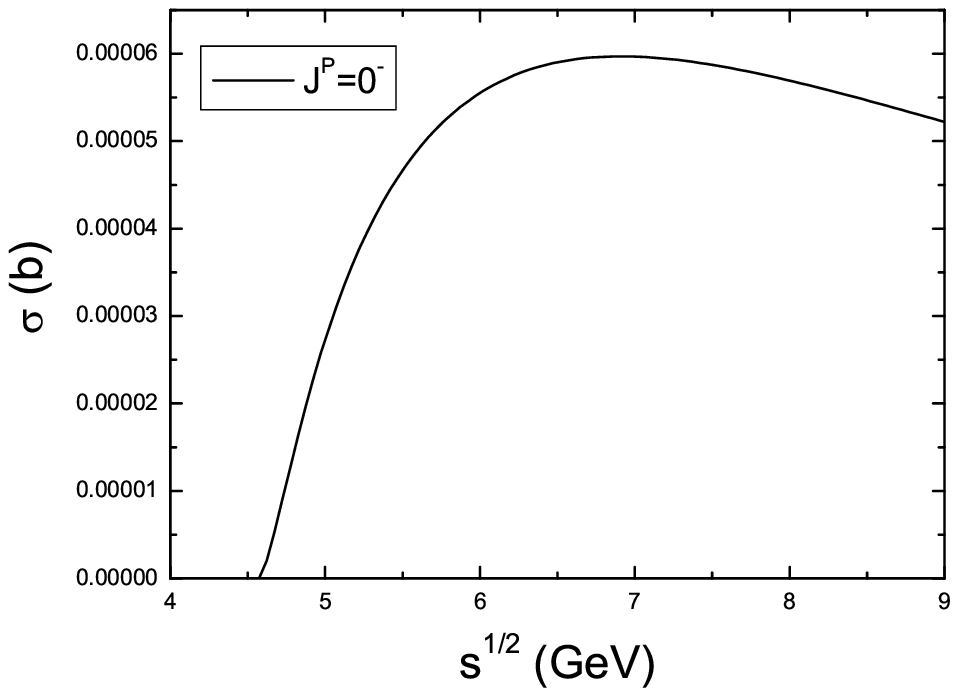}}&\scalebox{0.55}{\includegraphics{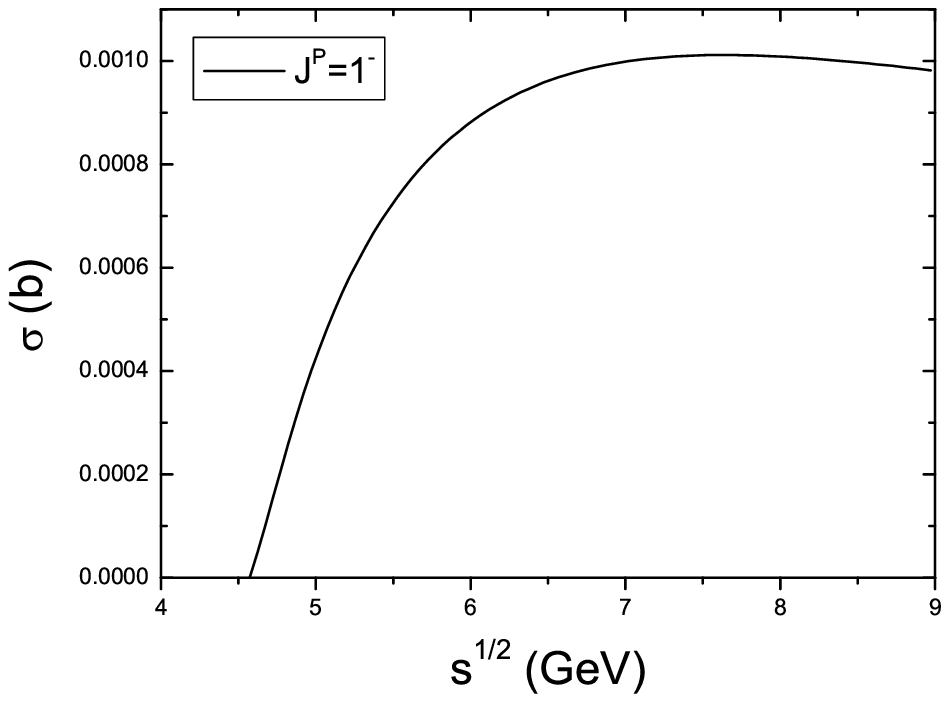}}
&\scalebox{0.55}{\includegraphics{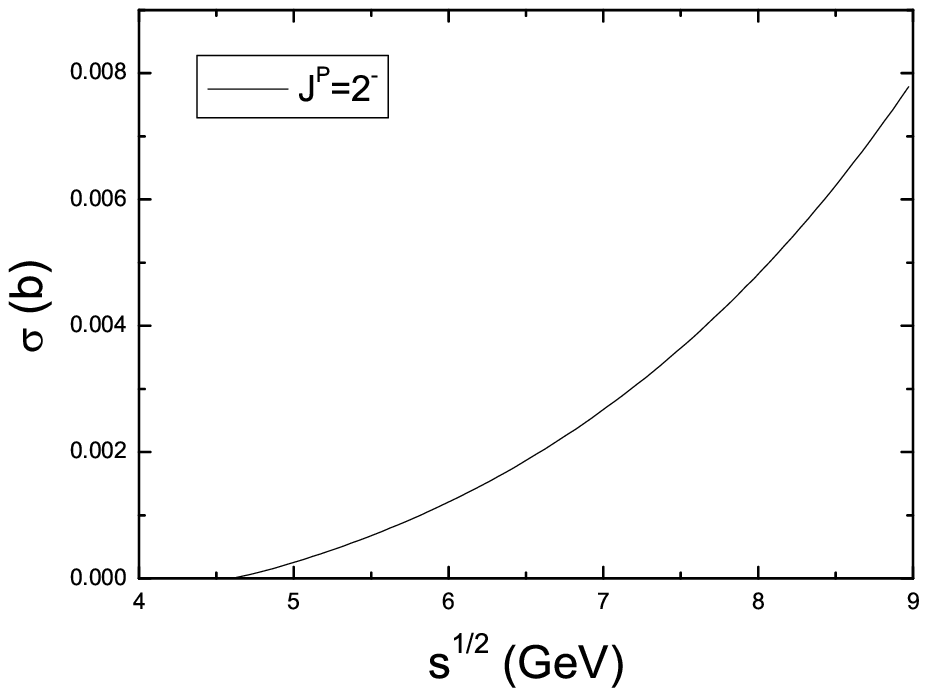}}\\
\end{tabular}
\caption{The cross section of $p\bar{p}\to Z^+(4430)\pi^-$ process
with three possible quantum numbers $J^{P}=0^-,1^-,2^-$ for
$Z^+(4430)$. \label{pp-Zpi}}
\end{figure}
\end{center}

\begin{center}
\begin{figure}[htb]
\begin{tabular}{ccccccc}
\scalebox{0.6}{\includegraphics{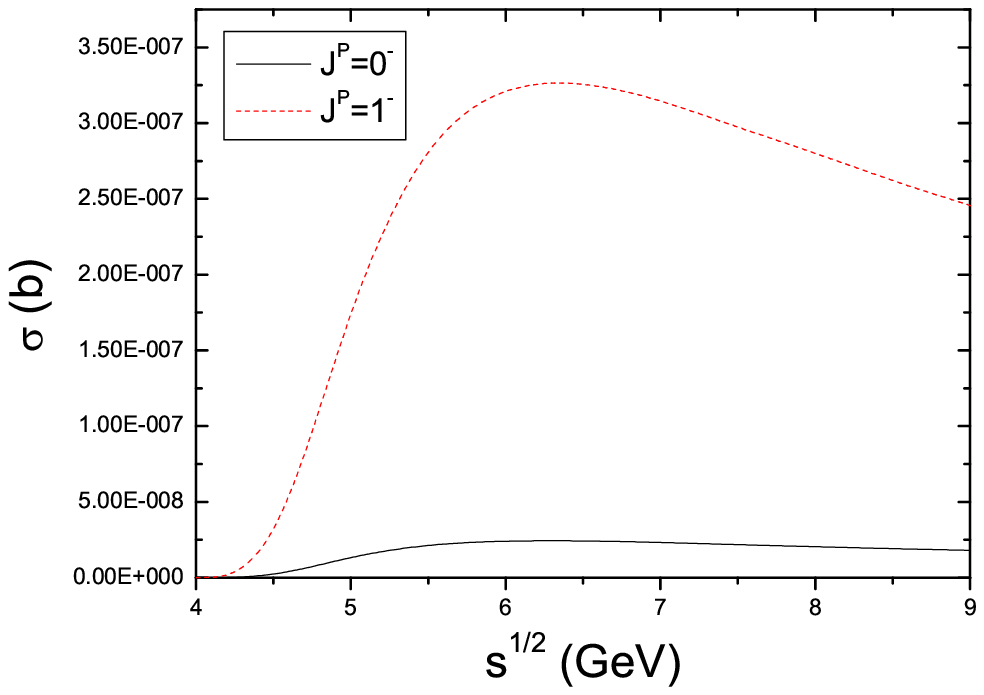}}&\scalebox{0.6}{\includegraphics{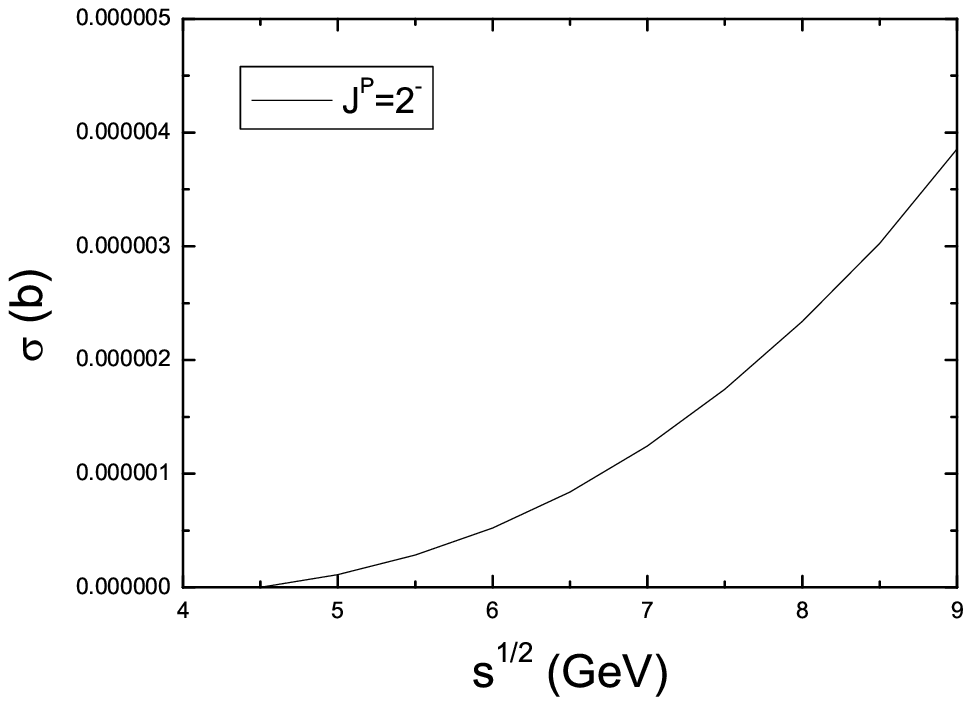}}\\
\end{tabular}
\caption{The cross section of $p\bar{p}\to \psi'\pi^+\pi^-$ with
the intermediate state $Z^+(4430)$. \label{pp-back}}
\end{figure}
\end{center}
\end{widetext}



It is difficult to obtain the value of the coupling constant
$g_{Z^+p\bar{n}}$ by a reliable dynamics calculation. On taking
the total width of $Z^+(4430)$ as the maximum of the
$\Gamma_{Z^+\to p\bar{n}}$, one gets the maximum of coupling
constant of $Z^+p\bar{n}$
\begin{eqnarray*}
&&g_{Zpn}^{(0)}=0.52,\quad g_{Zpn}^{(1)}=1.27,\quad
g_{Zpn}^{(2)}=0.36\;\mathrm{GeV}^{-2}.
\end{eqnarray*}
Using the above parameters, we obtain the dependence of the cross
section of $p\bar p\to Z^{+}(4430)\pi^-$ on the center-of-mass
energy $\sqrt{s}$, which is shown in Fig. \ref{pp-Zpi}.

The design luminosity of PANDA is $2\times 10^{32}$
cm$^{-2}s^{-1}$ for the collision of the antiproton and the proton
target \cite{panda}. Thus, the integrated luminosity is able to
reach approximately $1.5$ fb$^{-1}$/year if the machine is of
$50\%$ efficiency and in half a year of data taking \cite{ma}.
There exist sizable events for the production of $Z^+(4430)\pi^-$
in $p\bar p$ scattering, i.e. the maximum events with $\sqrt{s}=5$
GeV can be up to $8.9\times 10^{10}$, $1.4\times 10^{12}$ and
$2.9\times 10^{12}$, corresponding to $Z^+(4430)$ productions with
$J^{P}=0^-,1^-,2^-$, respectively. However, one needs to note that
the above observation is only valid on taking the total width of
$Z^+(4430)$ as the decay width of $Z^+(4430)\to p\bar n$. If the
order of magnitude of $\Gamma_{Z^{+}p\bar n}$ is about 1 MeV, the
above estimated events is suppressed by a factor $(1/44)^2\sim
5\times 10^{-4}$, still enough to comply with the needs of the
experimentalist. Even though $\Gamma_{Z^{+}p\bar n}$ is at the
order of keV, there also exists $10\sim 10^2$ events of
$Z^{+}(4430)$ production.

In Fig. \ref{pp-back}, we present the variation of the cross
section of $p\bar p\to Z^{+}(4430)\pi^-\to \psi'\pi^+\pi^-$ to
$\sqrt{s}$. Due to $p\bar p\to \psi'\pi^+\pi^-$ with intermediate
state $Z^+(4430)$ being a $2\to 3$ process, the cross section is
$3\sim 4$ order smaller than that of $p\bar p\to Z^+(4430)\pi^-$.
The line shape of the cross section of $p\bar p\to \pi^-
Z^+(4430)\to \psi'\pi^+\pi^-$ is similar to that of $p\bar p\to
Z^+(4430)\pi^-$.

Being a low-energy collision experiment of the antiproton with
fixed target, the PANDA experiment has unique advantages to search
the $Z^{\pm}(4430)$. Based on the above numerical analysis, we
suggest the forthcoming PANDA experiment to carry out the search
of $Z^\pm(4430)$, which will be helpful to finally establish
$Z^\pm(4430)$ and shed light on the nature of $Z^\pm(4430)$
further.

Figure \ref{pn-44} shows the cross sections of $\bar p n\to
Z^-(4430)\to \psi'\pi^-$ as the function of $\sqrt{s}$, which
includes the contribution from the background. Here we present the
different results by taking several values of the coupling
constant $g_{Z{p}n}$, which correspond to the
$\Gamma_{Z\bar{p}n}=44, 22 , 4.4$ and $1$ MeV respectively. The
line shape of the cross section of $\bar p n\to Z^-(4430)\to
\psi'\pi^-$ shows the enhancement at energy $\sqrt{s}=4.433$ GeV,
which is different from the case of $Z^+(4430)$ production at
$p\bar p$ scattering process.

In terms of our calculation, we find that the largest production
rate of $Z^{+}(4430)$ mainly comes from the $p \bar{n}$
annihilation process. Unfortunately, the PANDA experiment mainly
refers to the collision of proton and antiproton. Thus, it is
difficult to test our proposal of $Z^{+}(4430)$ production by $p
\bar{n}$ annihilation based on present high-energy experiments.
One of the difficulties is that it is not easy to provide and
control the source of the neutron, which is waiting to be solved
by clever experimentalists.






\begin{widetext}
\begin{center}
\begin{figure}[htb]
\begin{tabular}{ccccccc}
\scalebox{0.53}{\includegraphics{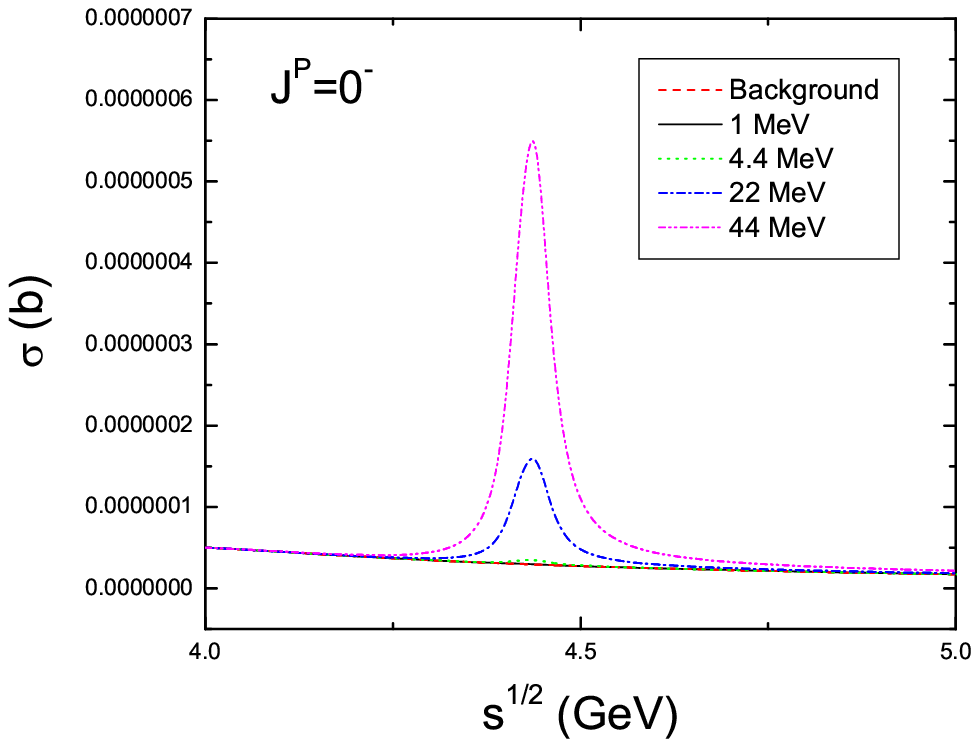}}&\scalebox{0.53}{\includegraphics{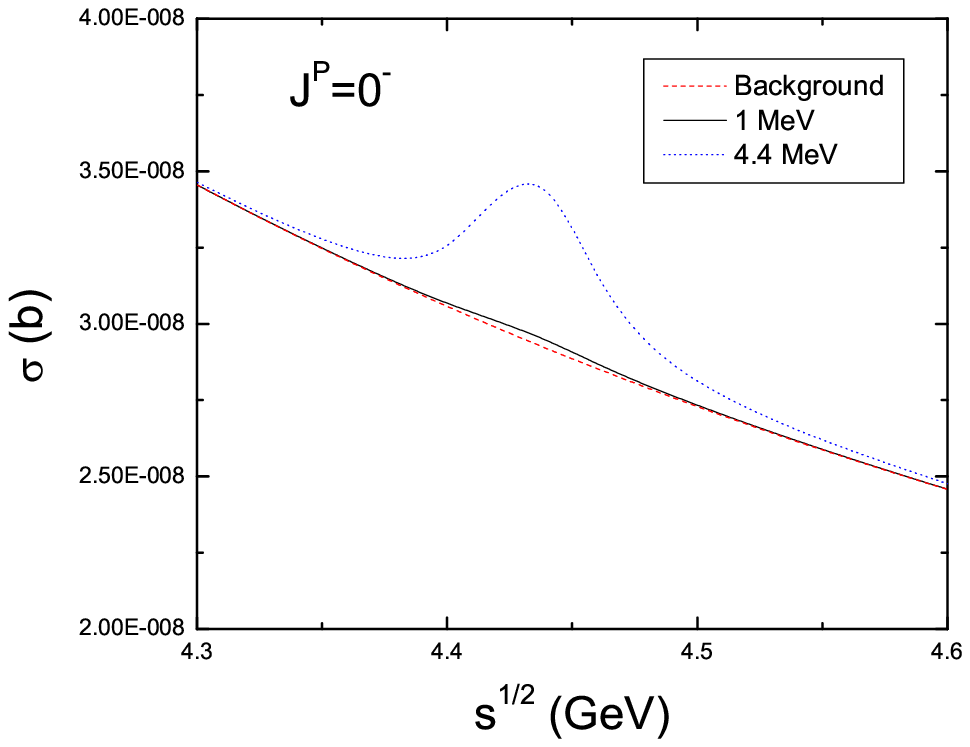}}&
\scalebox{0.53}{\includegraphics{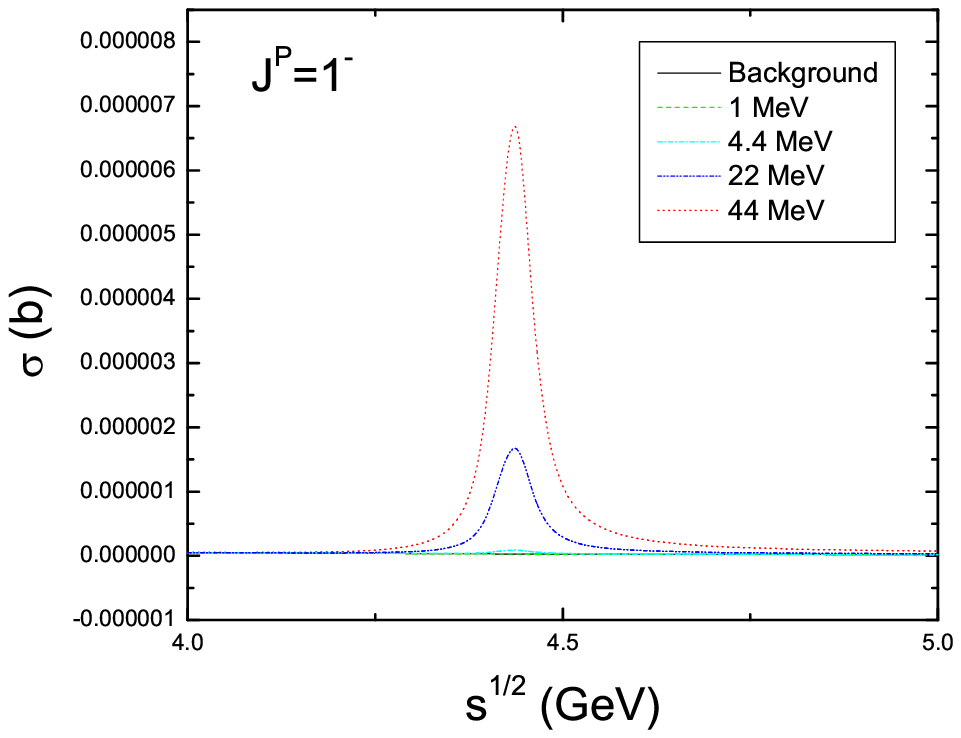}}\\\scalebox{0.53}{\includegraphics{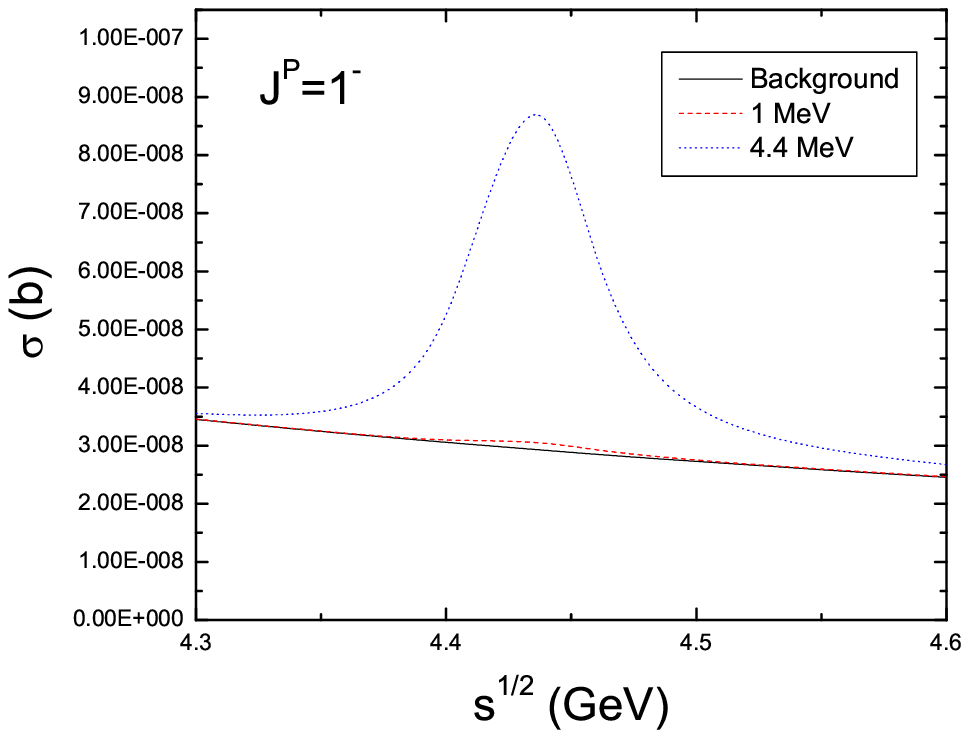}}&
\scalebox{0.53}{\includegraphics{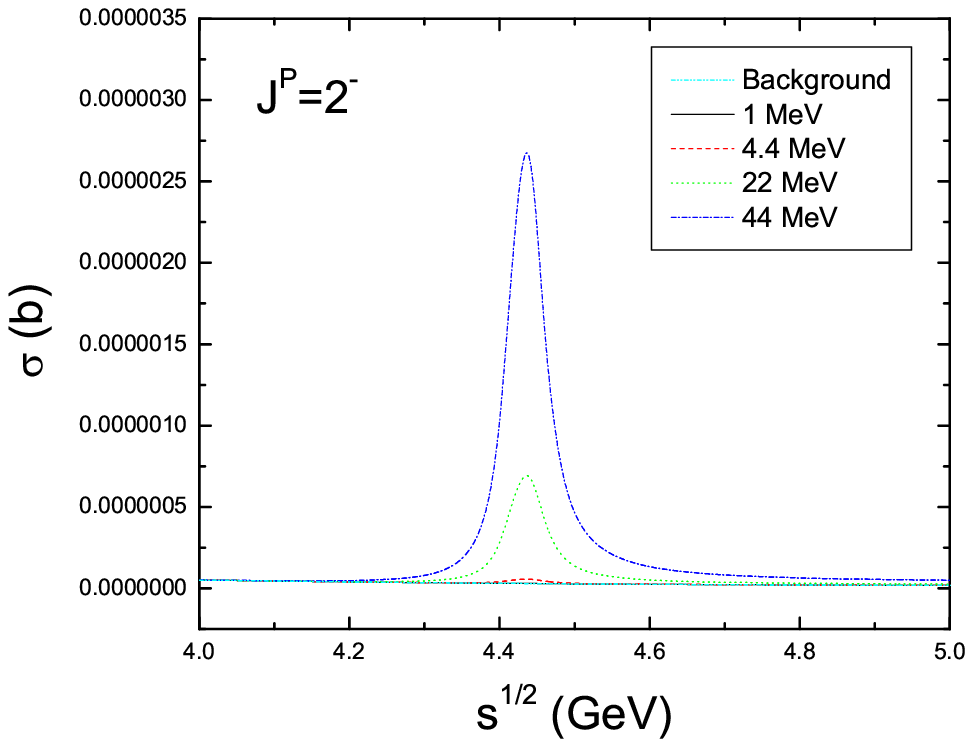}}&\scalebox{0.53}{\includegraphics{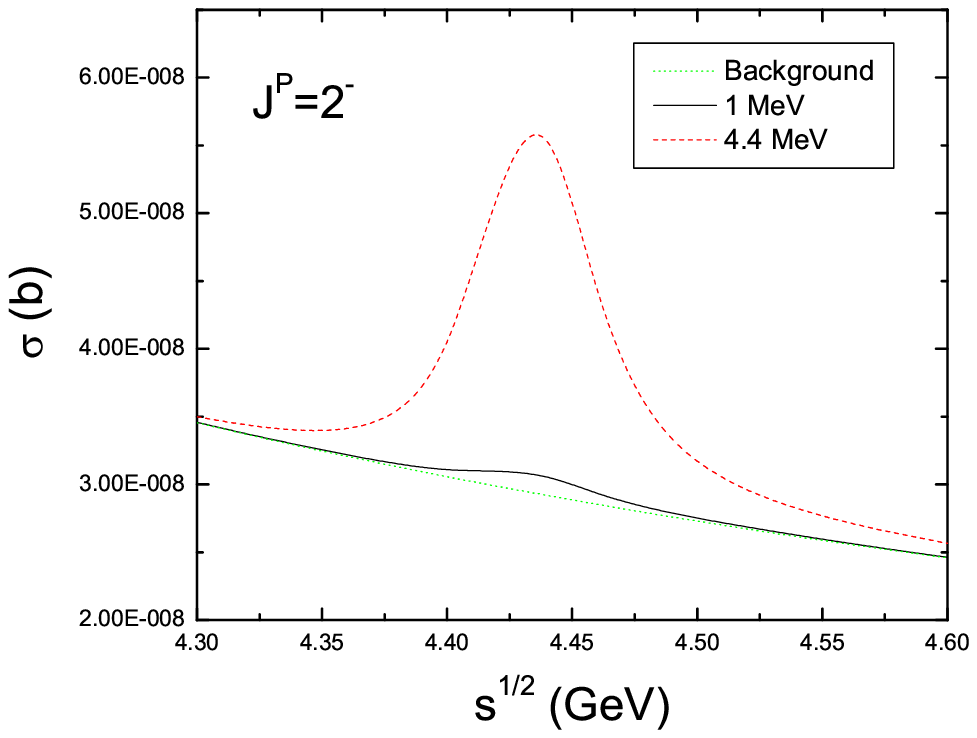}}\\
\end{tabular}
\caption{The cross section of the production of $Z^{-}(4430)$ in
$\bar{p}n$ annihilation process. \label{pn-44}}
\end{figure}
\end{center}
\end{widetext}

\vfill

\section*{Acknowledgments}

One of us (X.L.) would like to thank Prof. Eef van Beveren for
useful discussion and fruitful comments. This work was supported
by the \emph{Funda\c{c}\~{a}o para a Ci\^{e}ncia e a Tecnologia of
the Minist\'{e}rio da Ci\^{e}ncia, Tecnologia e Ensino Superior}
of Portugal (SFRH/BPD/34819/2007) and the National Natural Science
Foundation of China under Grant 10705001.



\begin{thebibliography}{99}
\bibitem{Belle-4430}Belle Collaboration, K. Abe et al.,
Phys. Rev. Lett. {\bf 100}, 142001 (2008).


\bibitem{rosner}J.L. Rosner, Phys. Rev. {\bf D 76}, 114002 (2007).
\bibitem{Meng}C. Meng and K.T. Chao, arXiv:0708.4222 [hep-ph].
\bibitem{xiangliu}X. Liu, Y.R. Liu, W.Z. Deng and S.L. Zhu, Phys. Rev. {\bf D 77}, 034003 (2008),
arXiv:0711.0494.
\bibitem{xiangliu-2}X. Liu, Y.R. Liu, W.Z. Deng, S.L.
Zhu, Phys. Rev. {\bf D 77}, 094015 (2008), arXiv:0803.1295
[hep-ph].
\bibitem{ding}G.J. Ding, arXiv:0711.1485 [hep-ph]; G.J. Ding, W.
Huang, J.F. Liu and M.L. Yan, arXiv:0805.3822 [hep-ph].

\bibitem{qsr}S.H. Lee, A. Mihara, F.S. Navarra and M. Nielsen,
arXiv:0710.1029 [hep-ph].
\bibitem{maiani}L. Maiani, A.D. Polosa and V. Riquer,
arXiv:0708.3997 [hep-ph].
\bibitem{cky}K. Cheung, W.Y. Keung and T.C. Yuan, Phys. Rev. {\bf D 76}, 117501 (2007).
\bibitem{Gershtein}S.S. Gershtein, A.K. Likhoded and G.P. Pronko,
arXiv:0709.2058 [hep-ph].
\bibitem{Bugg2}D.V. Bugg, arXiv:0802.0934 [hep-ph].
\bibitem{Qiao}C.F. Qiao, arXiv:0709.4066 [hep-ph].
\bibitem{braaten}E. Braaten and M. Lu, arXiv:0712.3885 [hep-ph].
\bibitem{lu}Y. Li, C.D. Lu and W. Wang, Phys. Rev. {\bf D 77}, 054001
(2008).
\bibitem{close}X.H. Liu, Q. Zhao and F.E. Close, arXiv:0802.2648
[hep-ph].

\bibitem{open}X. Liu, B. Zhang and S.L. Zhu, Phys. Rev. {\bf D 77}, 114021, (2008), arXiv:0803.4270
[hep-ph].

\bibitem{Matsuki}T. Matsuki, T. Morii and K. Sudoh,
arXiv:0805.2442 [hep-ph].

\bibitem{panda}PANDA Collaboration, Technical Progress Report
for: PANDA, http://www-panda.gsi.de.

\bibitem{barnes}T. Barnes and X. Li, Phys. Rev. {\bf D 75}, 054018
(2007).

\bibitem{ma}G.Y. Chen and J.P. Ma, arXiv:0802.2982 [hep-ph].


\bibitem{GMP}M.K. Gaillard, L. Maiani and R. Petronzio, Phys.
Lett. {\bf B 110}, 489 (1982).

\bibitem{LBW}A. Lundborg, T. Barnes and U. Wiedner, Phys. Rev. {\bf D
73}, 096003 (2006).

\bibitem{coupling-nnpi}Z. Lin, C.M. Ko and B. Zhang, Phys. Rev. {\bf C 61}, 024904 (2000); B. Holzenkamp et al. Nucl.Phys. {\bf A
500}, 485 (1989).

\bibitem{3-body}T. Hahm, \emph{FormCalc 5 User¡¯s Guide}, http://www.feynarts.de/formcalc.

\bibitem{PDG}W.M. Yao et al., Particle Data Group, J. Phys. G {\bf 33}, 1
(2006).

\end{thebibliography}
\end{document}